\documentclass[11pt,a4paper]{article}

\usepackage[affil-it]{authblk}
\usepackage[margin=3.0cm]{geometry}
\usepackage{amsmath,amssymb,amsfonts,upref}
\usepackage{amsthm}
\usepackage{mathtools}
\usepackage{tabularx}

\title{Cybernetical Concepts for Cellular Automaton and Artificial Neural Network Modelling and Implementation}
\author[1]{Patrik Christen\footnote{Corresponding author: patrik.christen@fhnw.ch. The authors contributed equally to this work.}}
\affil[1]{Institute for Information Systems, FHNW University of Applied Sciences and Arts Northwestern Switzerland, Riggenbachstrasse 16, 4600 Olten, Switzerland}
\author[2]{Olivier Del Fabbro}
\affil[2]{Chair for Philosophy, ETH Zurich, Clausiusstrasse 49, 8092 Zurich, Switzerland}
\date{{\small Appeared in \textit{2019 IEEE International Conference on Systems, Man and Cybernetics (SMC), 4124-4130, 2019.}}\\ \vspace{22pt}24 November 2019 (last revised 31 August 2020)}

\begin{document}

\maketitle

\abstract{As a discipline cybernetics has a long and rich history. In its first generation it not only had a worldwide span, in the area of computer modelling, for example, its proponents such as John von Neumann, Stanislaw Ulam, Warren McCulloch and Walter Pitts, also came up with models and methods such as cellular automata and artificial neural networks, which are still the foundation of most modern modelling approaches. At the same time, cybernetics also got the attention of philosophers, such as the Frenchman Gilbert Simondon, who made use of cybernetical concepts in order to establish a metaphysics and a natural philosophy of individuation, giving cybernetics thereby a philosophical interpretation, which he baptised allagmatic. In this paper, we emphasise this allagmatic theory by showing how Simondon's philosophical concepts can be used to formulate a generic computer model or metamodel for complex systems modelling and its implementation in program code, according to generic programming. We also present how the developed allagmatic metamodel is capable of building simple cellular automata and artificial neural networks.}



\newpage
\section{Introduction}

As scientific branch, cybernetics has a long and rich history. In its first generation of genesis it did not only gather famous and prodigy scholars such as John von Neumann, Stanislaw Ulam, Norbert Wiener, Warren McCulloch, Walter Pitts, and Claude Shannon, it also had a worldwide span emerging in other countries than the United States, such as Germany, France, and Russia \cite{c1,c2}. Moreover, as cybernetics grows, it also got the attention of other university disciplines such as philosophy: e.g. philosophy of biology \cite{c3}, psychoanalysis \cite{c4}, hermeneutics \cite{c5}, and pragmatism \cite{c6}.

What strikes most is the universal claim proponents of cybernetics make \cite{c7}. As Wiener puts it, cybernetics is born in-between classical scientific disciplines, e.g. in no man's land, and with its very general concepts, such as system or feedback, and with the help of its mathematical tools, it is apt to tackle every possible scientific and technological problem \cite{c8}. As the famous Macy conferences show, a broad range of subjects is covered including psychology, neurophysiology, information theory, quantum mechanics, sociology, and robotics \cite{c9}.

One of these numerous philosophers who followed the formation and evolution of cybernetics meticulously, is the Frenchman Gilbert Simondon (1924-1989) \cite{c10,c11}. However, for Simondon the universal claim of cybernetics is indeed important but not a prime issue. Rather, cybernetics comes up with new concepts, such as system or feedback, allowing to interpret real phenomena from a different perspective.

What then results in Simondon's work is a natural philosophy looking at all possible phenomena, e.g. physical, chemical, biological, technical, psychological, and social in order to describe these realms with the help of cybernetical concepts \cite{c12}. All kinds of phenomena become systems with feedback mechanisms communicating with other systems. But, methodologically, Simondon is not simply arguing with these concepts on an abstract metaphysical level. He goes on to describe all sorts of systems in a very meticulous fashion, reinterpreting scientific theories cybernetically. In other words: Simondon creates models of real phenomena, such as the wave-particle duality in quantum mechanics, crystals, zoological colonies of polyps, insects and mammals, human individuals and collectives, combustion engines and vacuum tubes. However, writing in the 1950's and 1960's, technological and scientific advancements of modelling tools were not as much progressed as today. Hence, Simondon's models are interpretations and descriptions done by himself with a piece of a paper and a pen writing plain text.

Nevertheless, there seems to be a strong analogy between Simondon's general concepts and his methodology and today's modelling techniques. This becomes especially apparent in models such as cellular automata, agent-based models, and artificial neural networks that originated in the works of cyberneticians such as von Neumann, Ulam, McCulloch, and Pitts. Moreover, researchers in complex systems, originating in Warren Weaver's famous paper, {\it Science and Complexity} \cite{c13} and today's proponents such as Brian Arthur \cite{c14}, Stuart Kauffman \cite{c15}, Stephen Wolfram \cite{c16} and many more being affiliated to the Santa Fe Institute, draw similar philosophical consequences from their works as Simondon. Particularly, concepts such as emergent behaviour, non-linearity, processual and open-ended computation, and self-organisation are as much important for Simondon as they are for complexity researchers \cite{c17,c18}.

Yet, in this paper we will not give a philosophical interpretation of today's computer models highlighting the analogies to Simondon's philosophy. Rather, we presuppose these analogies as postulate in order to formulate and implement in program code Simondon's general concepts, obtaining thereby a general computer model or metamodel for cybernetical or complex systems modelling. As Simondon started from the works of the first generation of cybernetics to methodologically re-interpret them, we in turn start from Simondon's general concepts and methods in order to implement them into contemporary computer models, i.e. cellular automata and artificial neural networks, obtaining thereby a new modelling method. Simondon calls his own method and theory allagmatic because he wants to show how systems are changing on a processual basis \cite{c19}. Allagmatic is derived from the Greek verb, {\it allatein}, meaning change, transition, transformation, and exchange. In reference to Simondon, we propose to call our method as well allagmatic, allowing us in turn to anticipate developments in the realm of computer modelling, particularly meta-modelling and meta-programming.

Lastly, in cultural studies and history of science and technology, cybernetics as institutional discipline has been proclaimed dead and vanquished \cite{c20}. This might be true, compared to disciplines such as physics or chemistry. However, with the present allagmatic meta-modelling and meta-programming, it is our intention to show that methods and concepts even of the early generation in cybernetics can still be used as guidance in today's computer models. We think that cybernetics is still very much at the heart of modern computer modelling.

In this paper we will therefore describe what Simondon's general concepts are and how we formulated and implemented them into general model building blocks of an allagmatic metamodel. Furthermore, we provide experiments of how simple cellular automata and artificial neural networks are created from the metamodel to test the feasibility.

\section{Gilbert Simondon's Philosophical Concepts Used In Computer Models}

Simondon uses cybernetic concepts in order to describe all sorts of phenomena and objects in reality. The very basic concepts he relies on are {\it structure} and {\it operation}, which in turn form together a {\it system}. In this section, we will not necessarily explain which role these concepts play in Simondon's cybernetical metaphysics, rather we will show how we translated and applied them to computer modelling of complex systems (Fig.~\ref{fig:Figure1}).

\subsection{Structure and Space}

Structure describes everything that has to do with the spatial configuration of the considered system such as agents, cells, nodes, lattices, and grids. Here, no processes and no dynamics are present. Structure represents the spatial dimension, the topological form of the system in its very basic configuration.

\subsection{Operation and Time}

Operation, on the contrary, represents the considered system's dynamic and processual, temporal dimension. It shows not only how the system behaves, it also shows how, e.g. cells, agents, or nodes implicate and influence each other reciprocally. On a very basic level it also defines how the structural dimension dilates topologically over time, that is how single cells, agents, or nodes are initially formed spatially. This can be compared to drawing a straight line on a piece of paper \cite{c19}. During the motion of drawing, the structural-topological dimension of the line is formed at the same time as the temporal operation of drawing is moving on. In short, operation defines how agents are created over time and once formed, how they behave in their specific milieu or environment.

\begin{figure}
  \includegraphics[width=\linewidth]{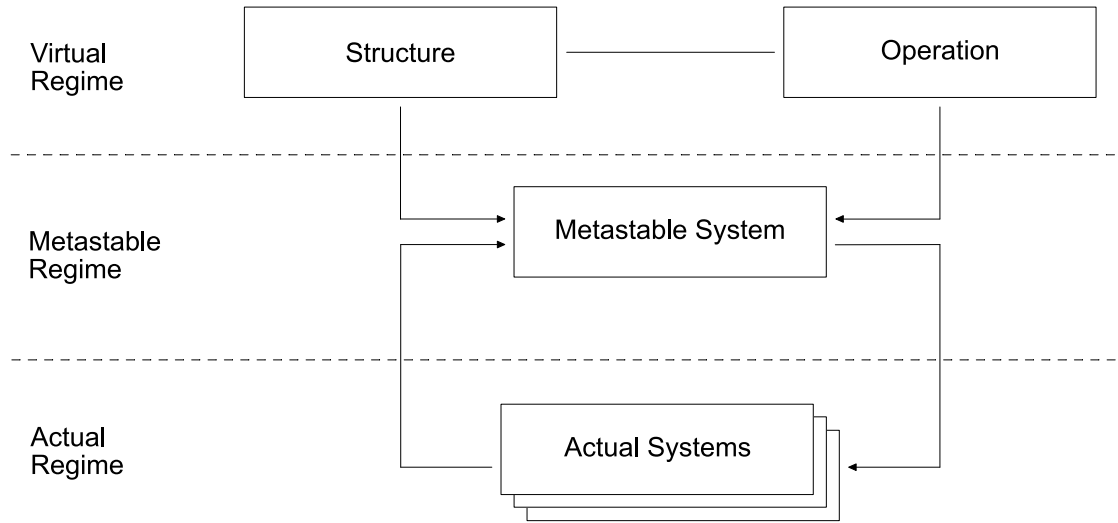}
  \caption{Philosophical concepts as proposed by the French Philosopher Gilbert Simondon and applied to computer modelling of complex systems.}
  \label{fig:Figure1}
\end{figure}

\subsection{Systems and Metastable Systems}

In actual reality, no system is constituted only by operations or structures solely, simply because every system has a spatial {\it and} a temporal dimension. Hence, a system is the product of structure and operation put together, without any parametrisation.

However, before becoming an actual system, when structure and operation are combined and parameterised, we need to introduce the concept of metastability. This means, that by combining structure and operation on a very basic level, several parameters such as initial conditions of the parameter states and the dynamical update function need to be defined in order to obtain a whole computing model. We name this system, where basic structures and operations are fed with this additional parameters, {\it metastable system}, because it represents a transition regime, which is partly virtual and partly actual. Hence, on a very basic level, structure and operation are defined in a virtual regime, whereas the actual model, including initial conditions and update functions computes in actuality (Fig.~\ref{fig:Figure1}).

The way this computation is processing, e.g. how structure and operation are interchanging and how they have causal effects on each other, is what Simondon calls allagmatic. Hence, allagmatic gives an account of how systems change spatially over time.

\section{Allagmatic Metamodel Formulation}

In the following, these philosophical concepts are built into general model building blocks of a computer model. These building blocks are fundamental to every computer model and as such they are independent from concrete models such as cellular automata and artificial neural networks.

\subsection{Model of Structure and Operation}

At the most abstract level, a computer model consists of at least one structure and at least one operation. They are described in the virtual regime and can be regarded as abstract descriptions of the spatial and temporal dimensions of a computer model (Fig.~\ref{fig:Figure1}).

Structure can be regarded as a $p$-tuple $e=\left(e_1,e_2,e_3,\dots,e_{p}\right)$ of $p$ basic {\it entities} such as cells, nodes, agents, or elements forming some kind of topology such as a lattice, grid, or network. The neighbourhood in a lattice or grid and the connections in a network can be regarded as a $q$-tuple $m$ consisting of $q$ neighbouring entities defined as the {\it milieu} of an entity. For the whole model or system, a milieu matrix $\mathbf{M}$ equivalent to the adjacency matrix for representing graphs \cite{c21} can be formulated containing all the information of how entities are spatially positioned. Structure, therefore, can be abstractly described by entities and their milieus with the tuple $e$ and the tuple $m$ defined for every $e_i$ or alternatively with the tuple $e$ and a matrix $\mathbf{M}$.

Operation, on the other hand, can be regarded as a process where entities $e$ belonging to the milieu $m$ are formed spatially and as a function $\phi$  that updates the set of potential states $s$ of the entities $e_i \in s$ over time. The state of the entity $e_i$ at the next discrete time step $t+1$ is thereby determined based on the states of its milieu entities $m_i$ and the state of itself at $t$. For a single entity $e_i$, operation can therefore be abstractly described by the {\it update function}
\begin{equation}
\phi:s^{q+1} \rightarrow s.
\end{equation}

\subsection{General Building Blocks of Model}

We therefore propose the model building blocks {\it entity}, {\it milieu}, and {\it update function} that together fully describe the general computer model on a virtual level. Notice, that these building blocks are described formally only on the virtual level without any actual values. Furthermore, they represent basic building blocks of a general computer model which can also consist of more than one structural and operational element.

\subsection{Model of Systems and Metastable Systems}

Hence, if the basic building blocks, as derived from structure and operation, are therefore combined, they can form a system. Feeding this system with parameters forms a metastable system that can finally act in the actual regime. The metastable system can thus be regarded as the initialised computer model, because it is here where all parameters are defined. Its execution is then occurring in the actual regime. The actual parameters consist of a specific update function $\phi$  including the concrete definition of $\mathbf{M}$ and update rules, boundary conditions that are described with $\phi$, and the initial conditions including the number of entities $q$ and their concrete set of possible states $s$.

Therefore, the two regimes, virtual and actual, are divided by a third interjacent regime of metastability (Fig.~\ref{fig:Figure1}). The formation of the metastable system, through the complementarity of structure and operation, reveals that the metastable system is the result of a superposition of structure and operation. Together with this superposition, the actual parameters at $t=0$, are defined arbitrarily or given by a particular application.

\section{Allagmatic Metamodel Implementation}

The formally described model building blocks entity, milieu, and update function describing the general computer model on an abstract level are in the following used to implement computer models with program code. Notice, that at this stage the specific type of model, such as cellular automaton or artificial neural network is not yet defined.

\subsection{Implementation of Structure}

Structure describing the spatial dimension is defined by the entities tuple $e$ and the respective milieus in matrix $\mathbf{M}$. Spatiality is generally well represented by data structures and data members in procedural and object-oriented programming, respectively. Mathematical tuples are possible to implement in program code with arrays or similar data containers. It follows that structure is programmed with an array of entities, where each array element contains an entity object. The milieu of each entity is described by a two-dimensional array representing the milieu matrix. Because structure is part of the virtual regime, the size of the arrays as well as their element data type should not be defined beforehand, since the application and the respective input data will define them.

We implemented structure and thus the entities $e$ with a vector template in C++. A class \texttt{Entity} with the member data \texttt{state} is defined. Template meta-programming of C++ \cite{c22} is used to define a generic type for \texttt{state}. In another class, the \texttt{System} class, a vector called \texttt{systemEntities} with element type \texttt{Entity} is defined as member data. Each element of \texttt{systemEntities} thus contains an \texttt{Entity object}. The vector template allows dynamic size of \texttt{systemEntities}. The type of \texttt{state} and the size of \texttt{systemEntities} are defined by the initial conditions. The milieu matrix $\mathbf{M}$ is implemented as a data member \texttt{milieus} of the class \texttt{System} and in the form of a two-dimensional vector. It is a square matrix of the same size as the vector \texttt{systemEntities}. If only the interactions between the entities are described, the type Boolean would be sufficient. If, however, interactions are weighted in the model, as in artificial neural networks, a floating-point number is required. As this depends on the concrete model, a dynamic and generic type for \texttt{milieus} was implemented in the present study, with the vector template and template meta-programming of C++ accounting therefore for any kind of model.

\subsection{Implementation of Operation}

Operation describing the temporal dimension is defined by the update function $\phi$. It defines how \texttt{systemEntities} changes its element's \texttt{state} over time and also how they change their structure over time. The change of state itself is application or problem specific and represented by the function body. The change of structure is model specific and represented by the function inputs $e^{(t)}$ and $\mathbf{M}^{(t)}$, and the function output $e^{(t+1)}$ at a time $t$. Mathematical functions are possible to implement in program code by functions and member methods in procedural and object-oriented programming, respectively.

We here implement model specific member methods for simple cellular automata in the class \texttt{CA} and artificial neural networks in the class \texttt{ANN} with generic data types using template meta-programming. Their implementation is explained in more detail in the next section.

\section{Simple Experiments}

Starting from the general model building blocks, we describe the implementation of a simple cellular automaton and artificial neural network in this section, respectively. A two-state one-dimensional cellular automaton and a multilayer feedforward artificial neural network are implemented guided by the allagmatic model, e.g. the respective model building blocks.

\subsection{Implementation of Cellular Automaton}

The basic building blocks of a general computer model are in the following concretised in a cellular automaton. This model can be characterised by a discrete cellular state space $\mathcal{L}$ on which the dynamics unfolds, a local value space $\Sigma$ defining the set of possible states a cell can assume, boundary conditions, and a dynamic rule $\phi$ defining the temporal behaviour \cite{c23}. We consider a two-state one-dimensional cellular automaton, which is made of identical cells $c$ and apply a periodic boundary condition \cite{c24}. In this simple cellular automaton, $\mathcal{L}$ is a one-dimensional lattice connected together at both ends forming a ring. The $i$-th cell of the lattice can have one of two possible states, $c_i \in \Sigma = \{ 0,1 \}$. The state of $c_i$ at time $t+1$ is determined by the states of its direct neighbours $c_{i-1}$ and $c_{i+1}$, and itself $c_i$ at time $t$ through the dynamic rule or local transition function
\begin{equation}
c_{i}^{(t+1)}=\phi(c_{i-1}^{(t)},c_{i}^{(t)},c_{i+1}^{(t)}).
\end{equation}
The dynamic rule updates the cellular states over time in accordance with a set of $n_{\Sigma}^{n_{in}}$ prescribed rules, where $n_{\Sigma}$ is the number of possible cell states and $n_{in}$ the number of input parameters. There are $2^3=8$ possible rules in a two-state cellular automaton with a neighbourhood defined by the nearest neighbours of a cell.

In this very basic cellular automaton, structure represents the fact that there is a lattice of cells and that spatially a further cell is situated at $t+1$ (in the next generation) below the three cells considered as input parameters of $\phi$ at $t$ if the lattices of each time step are stacked on top of each other. Operation represents the fact that within the considered part of the lattice consisting of three cells, each cell is initially formed and thereby directly linked to its adjacent neighbour cell and that these three cells taken together inform the state of the cell below in the next generation. As we have already mentioned earlier, only as a couple spatial structure and temporal operation form a metastable system. However, in order to obtain an actual system, in this case a real computing cellular automaton, further actual parameters are needed, such as initial conditions and a specific set of rules in the transition function.

The update function \texttt{updateFunction} is here implemented as a static member method in the class \texttt{CA}. It takes the arguments \texttt{systemEntities}, \texttt{milieus}, and \texttt{rules} in the form of references. These are all member data of the class \texttt{System}. In the \texttt{main} method, an \texttt{entities} tuple of 31 \texttt{Entity} objects with \texttt{states} of the type Boolean are generated with a member method \texttt{generateInitialEntities} of the \texttt{CA} class. Wolfram's rule 110 \cite{c16} is encoded in \texttt{rules} of type Boolean. Structures are therefore implemented of type Boolean representing the set of two possible states $\Sigma=\{0,1\}$ and the operation as a truth table of three input parameters representing the nearest neighbours according to the milieu matrix $\mathbf{M}$ determining one output according to the specified dynamic rules in \texttt{rules}.

Starting with the initial condition of 
\begin{equation}
e^{(t=0)}=0000000000000001000000000000000,
\end{equation}
and randomly assigned rules, the aim of this experiment is to find the specific entity configuration 
\begin{equation}
e^{\text{target}}=1101011001111101000000000000000, 
\end{equation}
which is searched iteratively. It is the output of Wolfram's rule 110 at $t=15$ that is computed with the rules:
\begin{equation}
\begin{aligned}
\phi(0,0,0) & = 0, \\ 
\phi(0,0,1)	&= 1, \\
\phi(0,1,0)	&= 1, \\
\phi(0,1,1)	&= 1, \\
\phi(1,0,0)	&= 0, \\
\phi(1,0,1)	&= 1, \\
\phi(1,1,0)	&= 1, \\
\phi(1,1,1)	&= 0.
\end{aligned}
\end{equation}

In each iteration, the rules are assigned randomly and the output after 15 iteration is compared to the target entities. As long as less than 90\% of the states are matching, new rules are generated, and the computation of the cellular automaton is performed again.

Performing the experiment several times revealed that in most of the cases, a solution is found after less than 1000 iterations or rules. In every case, the exact rule 110 was found and no alternative solutions were found.

Despite of the simplicity of this model, rule 110 is an exceptional and interesting cellular automaton update rule as it shows complex behaviour at the edge of order and chaos \cite{c15}. Furthermore, it is computationally universal, that is it can run any program or algorithm \cite{c25}.

\subsection{Implementation of Artificial Neural Network}

The basic building blocks of a general computer model are in the following concretised in a multilayer feedforward artificial neural network. Generally, this model can be characterised by input, hidden, and output layers of neurons also called perceptrons, where every neuron of a layer is linked with each neuron of the proceeding layer and from input to hidden to output layers. Each neuron, therefore, has a number of incoming activation signals $a_i$ from other neurons $i$ (or the initial conditions in the case of the input layer) and calculates an outgoing activation signal $a_j$ of neuron $j$ that will be the incoming activation signal for each of the neurons in the proceeding layer (or represent the result in case of the output layer). Each neuron $j$ contains an input function $in_{j}$ that calculates the weighted sum of the incoming activation signals $a_i$ as well as a transfer or activation function $g$ that calculates the outgoing activation signal $a_j=g(in_j)$ of the neuron $j$. Please note that if we assume a single-layer architecture, where three input neurons are directly linked to one output neuron, the weights $\omega$ and activation function $g$ of the artificial neural network can be chosen in such a way that, starting with the same initial condition $e^{(t=0)}$, the same result $e^{\text{target}}$ as with the cellular automaton from the previous experiment can be achieved. It follows that this kind of artificial network model could be well described with the present metamodel. However, in this experiment we implement a multilayer network architecture that is more widely used today.

The generic implementation of the building blocks entity and milieu allows to determine a multilayer feedforward artificial neural network by initialising the entities tuple \texttt{systemEntities} and the milieu matrix \texttt{milieus} with the respective values. Each neuron is represented by one \texttt{Entity} object and the member data \texttt{state} is initialised with a Boolean value to store the input and output values as well as the calculated output values of the hidden layers. The weights are stored in the milieu matrix \texttt{milieus}. Thus, the elements of \texttt{milieus} are initialised with a randomly generated floating-point number. Since \texttt{milieus} is an adjacency matrix, this allows storing the weights and topology of the network without the need of introducing an additional data structure, which is important with respect to the generality of the presented allagmatic metamodel. The update function \texttt{updateFunction} is here implemented as a static member method in the class \texttt{ANN}. It takes the arguments \texttt{systemEntities} and \texttt{milieus} in the form of references. These are all member data of the class \texttt{System}. It contains the input function $in_j$ and activation function $g$, traditionally used to calculate the activation from neurons $i$ to neuron $j$ in artificial neural networks. In our implementation, we use an input function that sums $n$ weighted incoming activation signals $a_i$ of neuron $j$ with
\begin{equation}
in_j=\sum^n_{i=0}\omega_{i,j}\cdot a_i,
\end{equation}
where $\omega_{i,j}$ are the weights from neurons $i$ to neuron $j$, and a threshold activation function
\begin{equation}
g(in_j)=
\begin{cases}
0 \text{ if } 0.5 > in_j\\
1 \text{ if } in_j \geq 0.5
\end{cases}
\end{equation}
and the perceptron learning rule
\begin{equation}
\omega_{i,j} \leftarrow \omega_{i,j} + r(y-a_j)a_i,
\end{equation}
where $r$ is the learning rate, $y$ the desired activation signal, $a_j$ the current outgoing activation signal, and $a_i$ the current incoming activation signals \cite{c26}.

In the \texttt{main} method, an \texttt{entities} tuple of 31 \texttt{Entity} objects with \texttt{states} of the type Boolean is generated with a member method \texttt{generateInitialEntities} of the \texttt{ANN} class, although the method is identical to the same method of the \texttt{CA} class. Each of the 15 layers in the multilayer network consists of 31 \texttt{Entity} objects. Note that neurons are only connected to the nearest three neurons in the previous layer and not to all of the neurons. As in the cellular automaton experiment, the initial conditions are
\begin{equation}
e^{(t=0)}=0000000000000001000000000000000,
\end{equation}
and the target entities are
\begin{equation}
e^{\text{target}}=1101011001111101000000000000000.
\end{equation}

Again, performing the experiment several times, it revealed that in most of the cases, a solution was found after less than 100\,000 iterations of building and running a new network from scratch. In none of the cases, the exact target value was reached. It stopped at the defined criterion of having matches of approximately 90\%.

\section{Conclusion and Outlook}

The purpose of the present study was to define general model building blocks for abstract computer models and to investigate how to implement them in program code. We addressed this with the help of Simondon's cybernetical metaphysics providing a guideline for defining model building blocks constituting a general computer model and implementing them with respective generic and dynamic programming. In this sense, it is not the intention of this paper to ontologically show what, e.g. complexity \cite{c27,c28} or a computer model \cite{c29,c30} is. Rather, philosophical concepts have been implemented directly into programme code \cite{c31,c32,c33}.

Model building blocks of a general computer model or metamodel were defined based on the concepts of structure and operation. Hence, a system is composed of some local elements that form the topology through connections. From this we revealed the model building blocks entity for these local elements and milieu for the connections between entities. Since structure is changing over time and the entity's state can be of different types, a dynamic and generic implementation of the data structure and type is required. While structure provides the spatial dimension for the operation to occur, operation, on the other hand, forms the evolving structure and defines the connections between entities. It thus guides the implementation by suggesting input and output parameters of the update function, which is the model building block of operation. Hence, structure and operation as philosophical concepts guide the definition of a general computer model and suggest meta-programming and object-oriented programming for its abstract implementation in the virtual regime. It needs to be stated that specific program implementations of cellular automata and artificial neural networks would be more efficient in terms of computing. However, we aimed for generality rather than efficiency. This generality allows us here to illuminate cybernetical concepts in modern computer models, but it also simplifies and encourages the programming of different computer models. It can therefore be used as an implementation guideline for programmers. In addition, since objects of models or systems are created and methods are available to manipulate them, the allagmatic metamodel could serve as basis for automatic model building where it would be possible to mix different model building blocks from different types of models \cite{c36}.

From the abstract definition of model building blocks, concrete models are created forming a so called metastable system. Transitioning the abstract model building blocks into a concrete computer model is achieved by feeding actual parameters into the metastable system. Here we show that these actual parameters can be regarded as concrete initial conditions and a specific update function sufficiently informing the metastable system.

It is important to highlight, that in the most abstract virtual regime, structure and operation are not hierarchically separated from each other. Therefore, every structure is operated, and every operation is structured simultaneously. Since this simultaneous action is successively unfolding into more concrete actual parameters, it is obvious that every actual system consists of both, structural-spatial and operational-temporal parameters. During computation, however, the virtual regime is not being significantly altered or fundamentally changed. It represents the most abstract form of structure and operation.

Due to this initial complementarity of structure and operation and its more concrete deployment in the metastable regime, the general description of our computer model is not a simple abstract and theoretical framework. Rather, we created an allagmatic method, which operates in-between abstract and concrete levels, computing constantly new models. Certainly, this new method has very abstract and general concepts borrowed from philosophy, but its power lies within the actual application and computation of these concepts, being able to create very concrete models. In this sense, it is rather a tool, i.e. a method than a theory.

In our experiments we chose concrete computer models of very different types in terms of structure and operation. In cellular automata, a specific spatial domain is discretised and represented with cells in the model, while the arrangement of neurons in artificial neural networks does not represent any specific spatial domain of the modelled system; except if neurons in a brain are represented. Also, operation does relate to some specific dynamic behaviour of the modelled system in cellular automata described by update rules whereas in artificial neural networks the input and activation functions are of statistical nature not representing any dynamics of the modelled system. Nevertheless, both models are here created from the present metamodel.

Being able to create computing cellular automata and artificial neural networks, makes the allagmatic method the abstract complement of already existing computer models. Hence, concrete models such as cellular automata and artificial neural networks can be created {\it within} the same framework of the allagmatic method. That is to say, on the basis between abstract and concrete computation, the allagmatic method is highly adaptable. Hence, the virtual regime is, except from its spatial and temporal constitution, not fixed to any specific type of computer model. Operating within the metastable regime and experimenting with the actual parameters allows to produce different types of models, such as cellular automata and artificial neural networks, if not also totally new types of models yet undiscovered.

Moreover, due to its abstractness it should therefore be possible to create models from already existing models of another type. For instance, it might be possible to create a cellular automaton from a neural network or vice versa. This would allow to observe how shared actual parameters are computed. We have seen, that starting from the virtual regime it is possible to create two different types of models, such as cellular automata and artificial neural networks. Even though these types may differ in the way they compute, both structure and operation are defined within the same set of generic parameters: e.g. initial conditions, possible states for entities, and update functions. If it will be possible to translate parameters from one model type to the other, it will also be possible to make models from models via the metamodel. That is to say, since cellular automata are at fist in the Peircean sense \cite{c38} iconic representations of the object, such as a drawing is representing directly the object it is supposed to depict, it will be possible to make simulations of artificial neural networks by using cellular automata. Since iconic and pictorial visualisations are easier to understand for humans, it could be possible to understand how artificial neural networks evolve and behave, e.g. by recognising complex patterns \cite{c37}. Such an endeavour can up until now only be formulated as a hypothesis requiring more experiments and research.

Similarly, since cellular automata are deterministic models \cite{c34}, that is to say, retrospectively their computation is traceable, and artificial neural networks are indeterministic, it could be possible to use the allagmatic method in the context of making artificial intelligence more interpretable or transparent\cite{c35} in another way. It might be possible to evolve cellular automaton rules to simulate the steps from one artificial neural network layer to the other. This would mean a translation of a purely statistical model into a model that represents the system that is modelled more pictorially or iconically, which can be regarded as a step towards an artificial intelligence more interpretable or transparent to humans.

Yet, by doing this implementation experiments it was primarily our intention to show that starting from philosophical concepts such as structure, operation, and system borrowed from Simondon's cybernetical metaphysics, building blocks of computer models can be built on an abstract level. Hence both, the abstract and the concrete definition provide guidance for mathematically describing and programmatically implementing computer models.

\section*{Acknowledgements}

This work was supported by the Hasler Foundation under Grant 18067.

\bibliographystyle{unsrt}
\bibliography{References_The_Allagmatic_Method.bib}

\begin{thebibliography}{10}

\bibitem{c1}
Slava Gerovitch.
\newblock {\em From {N}ewspeak to {C}yberspeak: {A} {H}istory of {S}oviet
  {C}ybernetics}.
\newblock The MIT Press, Cambridge and London, 2002.

\bibitem{c2}
Christopher~Charles Bissell.
\newblock Hermann {S}chmidt and {G}erman '{P}roto-{C}ybernetics'.
\newblock {\em Information, Communication \& Society}, 14(1):156--171, 2011.

\bibitem{c3}
Raymond Ruyer.
\newblock {\em La cybern\'etique et l'origine de l'information}.
\newblock Flammarion, Paris, 1954.

\bibitem{c4}
Jacque Lacan.
\newblock Psychanalyse et cybern\'etique, ou de la nature du langage.
\newblock In {\em Le moi dans la th\'eorie de Freud et la technique de la
  psychanalyse}, pages 403--421. \'Editions du Seuil, Paris, 1978.

\bibitem{c5}
Martin Heidegger.
\newblock Interview with {R}. {A}ugstein. {N}ur ein {G}ott kann uns retten.
\newblock {\em Spiegel}, 23, 1966.

\bibitem{c6}
Laura Moorhead.
\newblock Down the rabbit hole: Tracking the humanizing effect of {J}ohn
  {D}ewey's pragmatism on {N}orbert {W}iener.
\newblock {\em 2014 IEEE Conference on {N}orbert {W}iener in the 21st Century},
  2014.

\bibitem{c7}
Claus Pias.
\newblock Zeit der {K}ybernetik -- {E}ine {E}instimmung.
\newblock In Claus Pias, editor, {\em Kybernetik: The Macy Conferences
  1946-1953, Essays und Dokumente, Band II}, pages 9--41. Diaphanes, Z\"urich
  and Berlin, 2004.

\bibitem{c8}
Norbert Wiener.
\newblock {\em Cybernetics: Or Control and Communication in the Animal and the
  Machine}.
\newblock The MIT Press, Cambridge, 1961.

\bibitem{c9}
Claus Pias.
\newblock {\em Kybernetik: The Macy Conferences 1946-1953, Essays und
  Dokumente, Band I}.
\newblock Diaphanes, Z\"urich and Berlin, 2003.

\bibitem{c10}
Gilbert Simondon.
\newblock \'{E}pist\'emologie de la cybern\'etique.
\newblock In {\em Sur la philosophie, 1950-1980}, pages 177--199. PUF, Paris,
  2016.

\bibitem{c11}
Gilbert Simondon.
\newblock Cybern\'etique et philosophie.
\newblock In {\em Sur la philosophie, 1950-1980}, pages 35--68. PUF, Paris,
  2016.

\bibitem{c12}
Gilbert Simondon.
\newblock {\em L'individuation \`a la lumi\`ere des notions de forme et
  d'information}.
\newblock Editions J\'er\^ome Millon, Grenoble, 2013.

\bibitem{c13}
Warren Weaver.
\newblock Science and complexity.
\newblock {\em American Scientist}, 36(4):536--544, 1948.

\bibitem{c14}
W.~Brian Arthur.
\newblock {\em Complexity and the Economy}.
\newblock Oxford University Press, New York, 2014.

\bibitem{c15}
Stuart~A. Kauffman.
\newblock {\em The Origins of Order: Self-Organization and Selection in
  Evolution}.
\newblock Oxford University Press, New York, 1993.

\bibitem{c16}
Stephen Wolfram.
\newblock {\em A New Kind of Science}.
\newblock Wolfram Media, Champaign, 2002.

\bibitem{c17}
Melanie Mitchell.
\newblock {\em Complexity: A Guided Tour}.
\newblock Oxford University Press, New York, 2009.

\bibitem{c18}
John~H. Holland.
\newblock {\em Signals and Boundaries: Building Blocks for Complex Adaptive
  Systems}.
\newblock The MIT Press, Cambridge, 2012.

\bibitem{c19}
Gilbert Simondon.
\newblock Allagmatique.
\newblock In {\em L'individuation \`a la lumi\`ere des notions de forme et
  d'information}, pages 529--536. Editions J\'er\^ome Millon, Grenoble, 2013.

\bibitem{c20}
Michael Hagner.
\newblock Vom {A}ufstieg und {F}all der {K}ybernetik als
  {U}niversalwissenschaft.
\newblock In Michael Hagner and Erich H\"orl, editors, {\em Die Transformation
  des Humanen, Beitr\"age zur Kulturgeschichte der Kybernetik}, pages 36--71.
  Suhrkamp, Frankfurt am Main, 2008.

\bibitem{c21}
Thomas~H. Cormen, Charles~E. Leiserson, Ronald~L. Rivest, and Clifford Stein.
\newblock {\em Introduction to Algorithms}.
\newblock The MIT Press, Cambridge, 2009.

\bibitem{c22}
Andrei Alexandrescu.
\newblock {\em Modern C++ Design: Generic Programming and Design Patterns
  Applied}.
\newblock Addison-Wesley, Upper Saddle River, 2001.

\bibitem{c23}
Andrew Ilachinski.
\newblock {\em Cellular Automata: A Discrete Universe}.
\newblock World Scientific Publishing, Singapore, 2001.

\bibitem{c24}
Klaus Mainzer and Leon Chua.
\newblock {\em The Universe as Automaton: From Simplicity and Symmetry to
  Complexity}.
\newblock Springer, Berlin Heidelberg, 2012.

\bibitem{c25}
Matthew Cook.
\newblock Universality in elementary cellular automata.
\newblock {\em Complex Systems}, 15(1):1--40, 2004.

\bibitem{c26}
Stuart~J. Russel and Peter Norvig.
\newblock {\em Artificial Intelligence: A modern Approach}.
\newblock Prentice Hall, Upper Saddle River, 2010.

\bibitem{c27}
Daniel~C. Dennett.
\newblock Real patterns.
\newblock {\em Journal of Philosophy}, 88(1):27--51, 1991.

\bibitem{c28}
James Ladyman, James Lambert, and Karoline Wiesner.
\newblock What is a complex system?
\newblock {\em European Journal for Philosophy of Science}, 3(1):33--67, 2013.

\bibitem{c29}
Frank Varenne.
\newblock {\em From Models to Simulations}.
\newblock Routledge, London, 2019.

\bibitem{c30}
Juan~M. Dur\'an.
\newblock {\em Computer Simulations in Science and Engineering:
  Concepts--Practices--Perspectives}.
\newblock Springer, Cham, 2018.

\bibitem{c31}
Katherine Munn and Barry Smith.
\newblock {\em Applied Ontology: An Introduction}.
\newblock Ontos Verlag, Heusenstamm, 2008.

\bibitem{c32}
Bernhard Ganter and Rudolf Wille.
\newblock {\em Formal Concept Analysis: Mathematical Foundations}.
\newblock Springer, Berlin and Heidelberg, 1999.

\bibitem{c33}
V.~K. Finn.
\newblock Epistemological foundations of the {JSM} method for automatic
  hypothesis generation.
\newblock {\em Automatic Documentation and Mathematical Linguistics},
  48(2):96--148, 2014.

\bibitem{c36}
Patrik Christen and Olivier Del~Fabbro.
\newblock Automatic programming of cellular automata and artificial neural
  networks guided by philosophy.
\newblock In Rolf Dornberger, editor, {\em New Trends in Business Information
  Systems and Technology: Digital Innovation and Digital Business
  Transformation}, pages 131--146. Springer, Cham, 2020.
\newblock arXiv:1905.04232.

\bibitem{c38}
Charles~S. Peirce.
\newblock What is a sign?
\newblock In Peirce~Edition Project, editor, {\em The Essential Peirce,
  Selected Philosophical Writings, Volume 2 (1893-1913)}. Indiana University
  Press, Bloomington and Indianapolis, 1998.

\bibitem{c37}
Stephen Wolfram.
\newblock Cellular automata as models of complexity.
\newblock {\em Nature}, 311(5985):419--424, 1984.

\bibitem{c34}
Stephen Wolfram.
\newblock 20 problems in the theory of cellular automata.
\newblock {\em Physica Scripta}, T9:170--183, 1985.

\bibitem{c35}
Paul Voosen.
\newblock The {AI} detectives.
\newblock {\em Science}, 357(6346):22--27, 2017.

\end{thebibliography}

\end{document}